\documentclass[twocolumn, pra, showpacs,superscriptaddress,floatfix]{revtex4}
\usepackage{graphicx}
\usepackage{dcolumn}
\usepackage{bm}
\usepackage{amsmath}

\begin{document}

\title{Ground-state properties of interacting two-component Bose gases in a hard-wall trap}
\author{Yajiang Hao}
\affiliation{Beijing National Laboratory for Condensed Matter
Physics, Institute of Physics, Chinese Academy of Sciences, Beijing
100190, China} \affiliation{Department of Physics, University of
Science and Technology Beijing, Beijing 100083, China}

\author{Yunbo Zhang}
\affiliation{Department of Physics and Institute of Theoretical
Physics, Shanxi University, Taiyuan 030006, China}

\author{Xi-Wen Guan}
\affiliation{Department of Theoretical Physics, Research School of
Physical Sciences and Engineering, Australian National University,
Canberra ACT 0200, Australia}

\author{Shu Chen}
\email{schen@aphy.iphy.ac.cn} \affiliation{Beijing National
Laboratory for Condensed Matter Physics, Institute of Physics,
Chinese Academy of Sciences, Beijing 100190, China}
\date{ \today}

\begin{abstract}
We investigate ground-state properties of interacting two-component
Bose gases in a hard-wall trap using both the Bethe ansatz and exact
numerical diagonalization method. For equal intra- and inter-atomic
interaction, the system is exactly solvable. Thus the exact ground
state wavefunction and density distributions for the whole
interacting regime can be obtained from the Bethe ansatz solutions.
Since the ground state is a degenerate state with total spin
$S=N/2$, the total density distribution are same for each degenerate
state. The total density distribution evolves from a Gauss-like Bose
distribution to a Fermi-like one as the repulsive interaction
increases. The distribution of each component is $N_\alpha/N$ of the
total density distribution. This is approximately true even in the
experimental situation. In addition the numerical results show that
with the increase of interspecies interaction the distributions of
two Tonks-Girardeau gases exhibit composite fermionization crossover
with each component developing N peaks in the strongly interacting
regime.
\end{abstract}
\pacs{67.85.-d, 67.60.Bc, 03.75.Mn}
 \maketitle


\section{Introduction}

Since two-component Bose-Einstein condensates (BECs) of trapped
alkali atomic clouds were realized experimentally
\cite{Myatt,Erhard}, low-dimensional multi-component Bose gases have
attracted much attention from theory and experiment due to their
connection to many areas of physics. Theoretical investigation
mainly focuses on the stability, phase separation, collective
excitation, Josephson-type oscillations and other macroscopic
quantum many-body phenomena \cite{Ho96,Ao,Pu,Cazalilla,Williams} in
the frame work of mean field theory. Other fundamental problems such
as topological defects, symmetry breaking effects also attract
growing interest \cite {symmetry,Timmermans}.

Advances in experiment with ultracold atoms provide exciting
opportunities to control and manipulate ultracold atom gases in
one-dimensional (1D) waveguides by tightly confining the atomic
cloud in two radial directions and weakly confining it along the
axial direction \cite {Stoeferle,Paredes,Toshiya}. Successful
realization of 1D interacting quantum degenerate gases enables us to
study novel many-body effect in various of interacting regimes, for
example, in the strongly interacting limit, i.e. the Tonks-Girardeau
(TG) gases \cite{Girardeau}. Tunability of the scattering length
cross Feshbach resonance allows experimentalists access to whole
interaction regime from a weakly interacting limit to a strongly
interacting limit. Strong correlation effect in 1D quantum
degenerate gases \cite{Olshanii} have been extensively studied in
recent years \cite{Olshanii2,Petrov,Dunjko,Chen,delcampo}. It is
shown that 1D quantum systems exhibit particular features which are
significantly different from its three dimensional counterpart.

The exact results for single component bosons with a repulsive
$\delta$-function interaction show that the density profiles evolve
from Gaussian-like distribution of Bosons to shell-structured
distribution of Fermions when interaction strength increases \cite
{Hao06,Hao07,Deuretzbacher,Zoellner,Cederbaum,Yin}. In order to
study the system in the strong interacting regime, non-perturbation
method is highly desirable and reliable because the mean field
theory is proven to be insufficient. For the multi-component quantum
gases, studies were carried out by means of various schemes such as
mean field theory \cite {Ho96,Ao,Pu,Cazalilla}, extended Bose-Fermi
mapping in the infinitely repulsive limit
\cite{Girardeau07,DeuretzbacherPRL,Guan,Hao08,Zoellner08} and exact
Bethe ansatz \cite{LiYQ,Gusj,Guan07,Fuchs}. The 1D homogeneous
two-component Bose gas with spin-independent $s$-wave scattering
(equal inter- and intra-species interacting strengthes) is exactly
solvable by the Bethe ansatz method in a whole physical regime
\cite{LiYQ,Gusj}. The ferromagnetic ordering in spinor Bose gases
was predicted some years ago \cite{Eisenberg,Yang}. However, the
system is no longer integrable if the atomic gas no longer has the
$SU(2)$ symmetry or it is trapped in an inhomogeneous potential. In
this situation numerical methods have to be exerted and rich
phenomena are shown for various parameters \cite {Zoellner08,Hao08}.

Previous study on the integrable two-component Bose gas mainly focuses on
the energy spectrum and excitation properties \cite{LiYQ,Guan07,Gusj,Fuchs}.
However, important quantities, which are related to the wave function of the
system and are accessible experimentally, such as the density distribution
and momentum distribution, are rarely addressed except in the limit of
infinitely repulsive interaction \cite{Girardeau07}. In this paper, we are
aimed to study the two-component bosonic systems with $SU(2)$ symmetry in
the whole interacting regime by means of the Bethe ansatz. In the case of
broken $SU(2)$ symmetry, we resort to exact diagonalization method. The
total density distribution and the density distribution for each component
can be derived from exact ground state wave function. In addition, numerical
method will be used to evaluate the reduced one-body density matrix and
momentum distribution of each component as well as inter- and intra-species
density-density correlations.

The present paper is organized as follows. Section II introduces the model
and gives the exact solutions by means of the Bethe ansatz for the
integrable point. In Section III the numerical diagonalization method is
introduced and we investigate the system for accessible experimental
parameters after checking the accuracy of the numerical result. Section IV
is devoted to the interaction of two TG gases. A summary is given in the
last section.

\section{Exact solution of two-component Bose gas}

We consider two-component Bose gas confined in a 1D hard wall trap of length
$L$, which is composed of two internal states (pseudospin $\left| \uparrow
\right\rangle $ and $\left| \downarrow \right\rangle $ denote state 1 and 2,
respectively) of the same kind of Bose atoms with equal mass $m_1=m_2=m$.
The atom numbers in each component are $N_1$ and $N_2$ and $N=N_1+N_2$ the
total number. The many body system can be described by the second quantized
Hamiltonian
\begin{eqnarray*}
\mathcal{H} &=&\int dx\sum_{\alpha =1,2}\left\{ \frac{\hbar ^2}{2m_\alpha }
\frac{\partial \hat{\Psi}_\alpha ^{\dagger }(x)}{\partial x}\frac{\partial
\hat{\Psi} _\alpha (x) }{\partial x}\right. \\
&&\left. +\frac{g_\alpha }2 \hat{\Psi}_\alpha ^{\dagger }(x) \hat{\Psi}%
_\alpha ^{\dagger }(x) \hat{\Psi}_\alpha (x) \hat{\Psi}_\alpha (x)\right\} \\
&&+g_{12}\int dx \hat{\Psi}_1^{\dagger }(x) \hat{\Psi}_2^{\dagger }(x) \hat{%
\Psi}_2(x) \hat{\Psi}_1(x),
\end{eqnarray*}
where $g_\alpha $ ($\alpha =1,2$) and $g_{12}$ denote the effective intra-
and inter-species interaction which can be controlled experimentally by
tuning the corresponding scattering lengthes $a_1$, $a_2$ and $a_{12}$,
respectively. The field operator $\Psi _\alpha ^{\dagger }(x)$ ($\Psi
_\alpha (x)$) creates (annihilates) an $\alpha $-component boson at the
position $x$. A standard rescaling procedure brings the Hamiltonian into a
dimensionless one
\begin{eqnarray}
\mathcal{H} &=&\int dx\sum_{\alpha} \left\{ \hat{\Psi}_\alpha ^{\dagger }
(x) \left[ -\frac{\partial^2} {\partial x^2} + U_\alpha \hat{\Psi}_\alpha
^{\dagger } \hat{\Psi}_\alpha \right] \hat{\Psi}_\alpha (x)\right\}
\nonumber \\
&&+2 U_{12}\int dx\hat{\Psi}_1^{\dagger }(x)\hat{\Psi}_2^{\dagger }(x) \hat{%
\Psi}_2(x) \hat{\Psi}_1(x)  \label{Ham}
\end{eqnarray}
with $U_\alpha =mg_\alpha /\hbar ^2$ ($\alpha =1,2,12$). Here we
have rescaled the energy and length in units of $\hbar ^2/2mL^2$ and
$L$. The system with equal interaction constants $U_1=U_2=U_{12}=c$
is integrable in both the periodic boundary \cite{LiYQ} and open
boundary conditions \cite{Gusj,Guan07} and the eigen problem for the
original Hamiltonian is reduced to solving the coordinate nonlinear
Schr\"{o}dinger equation
\begin{equation}
H \Psi \left( x_1,\cdots ,x_N\right) = E \Psi \left( x_1,\cdots ,x_N\right)
\end{equation}
with
\begin{equation}
H=-\sum_{j=1}^N\frac{\partial ^2}{\partial x_j^2}+2c\sum_{j<l}\delta
(x_j-x_l).
\end{equation}
The Hamiltonian $H$ commutes with the total spin operator $\hat{S}$,
{\it i.e.}, commutes with all the three components of the total spin
operator, so that they share a common set of eigenstates and the
system possesses a global SU(2) symmetry. Explicitly, the three
components of total spin operator are defined as
\[
\hat{S}^{\alpha}=\frac 12 \int dx  \sum_{\mu,\nu}
\hat{\Psi}_{\mu}^{\dagger }(x) \sigma^{\alpha}_{\mu,\nu}\hat{\Psi}
_{\nu}(x) ,
\]
where $\mu, \nu \in  \{\uparrow, \downarrow \}$ with $\uparrow$ ($
\downarrow$) corresponding to the $1$th component ($2$th component)
and $\sigma^{\alpha}$ ($\alpha=x,y,z$) denotes the Pauli matrices.

The coordinate wave function can be determined by means of Bethe
ansatz method and takes the following general form
\begin{eqnarray}
&&\Psi \left( x_1,\cdots ,x_N\right)  \nonumber \\
&=&\sum_{P,Q}\theta \left( x_{q_N}-x_{q_{N-1}}\right) \cdots \theta \left(
x_{q_2}-x_{q_1}\right) \times  \nonumber \\
&& \sum_{r_1,\ldots ,r_N}\left[ A\left( Q,rP\right) \exp \left(
i\sum_jr_{p_j}k_{p_j}x_{q_j}\right) \right] ,  \label{wavefunction}
\end{eqnarray}
where $Q=(q_1,q_2,\cdots ,q_N)$ and $P=(p_1,p_2,\cdots ,p_N)$ are
one of the permutations of $1,\cdots ,N$, respectively,  $A\left(
Q,rP\right)$ is the abbreviation of the coefficient $
A\left(q_1,q_2,\cdots,q_N;r_{p_1}p_1,r_{p_2}p_2,\cdots,r_{p_N}p_N\right)
$ to be determined self-consistently, and the summation $\sum_P$ ($
\sum_Q$) is done for all of them. Here $r_j=\pm $ indicates that the
particles move toward the right or the left, $\theta (x-y)$ is the
step function and the parameters $\left\{ k_j\right\} $ are known as
quasi-momenta. Moreover the wavefunction Eq. (\ref{wavefunction})
should fulfill the the open boundary condition \cite{Gaudin} for
hard wall trap
\[
\Psi \left( \cdots ,x_j=0,\cdots \right) =\Psi \left( \cdots
,x_j=L,\cdots \right) =0,
\]
which enforces the relations
\[
A\left(Q;\cdots,-p_i,\cdots,r_{p_N}p_N\right) = - A\left(Q;\cdots,
p_i,\cdots,r_{p_N}p_N\right) .
\]
Furthermore, the coefficients fulfill the general relations
\[
A(Q;\cdots i, j \cdots )=Y_{ji}^{ab}A(Q;\cdots j, i \cdots )
\]
with $Y_{ji}^{ab}$ given by
\[
Y_{ji}^{ab}=\frac{ (k_{j}- k_{i})P_{q_a q_b}- i c}{ k_{j}- k_{i} + i
c},
\]
where $P_{q_a q_b}$ is the permutation operator on $A(\cdots
q_a,q_b,\cdots;P)$ so that it interchanges $q_a$ and $q_b$
\cite{Gusj,LiYQ,Guan07,YangCN}. For the eigenstate with total spin
$S=N/2-M$ ($0\leq M \leq N/2$), the Bethe ansatz equations
\cite{Gusj,Guan07} satisfied by the quasi-momentum \{$k_j$\} and
spin rapidity \{$\Lambda _\alpha $\} are given by
\begin{eqnarray*}
\exp(2ik_jL) &=&\prod_{l=1\neq j}^N \left(\frac{k_j-k_l+ic}{k_j-k_l-ic}\frac{%
k_j+k_l+ic}{k_j+k_l-ic}\right) \times \\
&& \prod_{\alpha =1}^{M}\left(\frac{k_j-\Lambda _\alpha -ic^{\prime
}}{k_j-\Lambda _\alpha +ic^{\prime }}\frac{k_j+\Lambda _\alpha
-ic^{\prime }}{k_j+\Lambda _\alpha +ic^{\prime }}\right),
\end{eqnarray*}
\begin{eqnarray*}
&&\prod_{l=1}^N\left(\frac{\Lambda _\alpha -k_l-ic^{\prime
}}{\Lambda _\alpha -k_l+ic^{\prime }}\frac{\Lambda _\alpha
+k_l-ic^{\prime }}{\Lambda _\alpha
+k_l+ic^{\prime }}\right) \\
&=&\prod_{\beta \neq \alpha }^{M}\left(\frac{\Lambda _\alpha -\Lambda _\beta -ic}{%
\Lambda _\alpha -\Lambda _\beta +ic}\frac{\Lambda _\alpha +\Lambda _\beta -ic%
}{\Lambda _\alpha +\Lambda _\beta +ic}\right),
\end{eqnarray*}
with $c^{\prime}=c/2$. The energy eigenvalue is $E=\sum_{j=1}^Nk_j^2$.
Taking the logarithm of Bethe ansatz equations, we have
\begin{eqnarray}
k_jL &=&\pi I_j -\sum_{l=1}^N\left( \tan ^{-1}\frac{k_j-k_l}c+\tan ^{-1}%
\frac{k_j+k_l}c\right)  \nonumber \\
&& +\sum_{\alpha =1}^{M}\left( \tan ^{-1}\frac{k_j-\Lambda _\alpha }{%
c^{\prime }}+\tan ^{-1}\frac{k_j+\Lambda _\alpha }{c^{\prime }}\right) ,
\label{BAE}
\end{eqnarray}
\begin{eqnarray}
&&\sum_{j=1}^N \left( \tan ^{-1}\frac{\Lambda _\alpha -k_j}{c^{\prime }}%
+\tan ^{-1}\frac{\Lambda _\alpha +k_j}{c^{\prime }}\right)  \nonumber \\
&=&\pi J_\alpha +\sum_{\beta \neq \alpha }^{M}\left( \tan ^{-1}\frac{\Lambda
_\alpha -\Lambda _\beta }c+\tan ^{-1}\frac{\Lambda _\alpha +\Lambda _\beta }%
c\right) .  \nonumber
\end{eqnarray}
Here the quantum numbers $I_j$ and $J_\alpha $ take integer or
half-integer values, depending on whether $N-M$ is odd or even. The
ground state corresponds to the case with $M=0$
\cite{LiYQ,Gusj,Guan07,Fuchs}. For the ground state $I_j=\left(
N+1\right) /2-j$ and $J_\alpha $ is an empty set, and the Bethe
ansatz equations reduce to the situation of Lieb-Liniger Bose gas.
In this case, the coefficient $A(Q;p_1,p_2,\cdots,p_N)$ can be
explicitly expressed as
\begin{eqnarray*}
& & A(Q;p_1,p_2,\cdots,p_N) \\
&=& (-1)^P
\prod_{j<l}^{N}(ik_{p_l}-ik_{p_j}+c)(ik_{p_l}+ik_{p_j}+c),
\end{eqnarray*}
with $(-1)^P= \pm 1$ denoting sign factors associated with even or
odd permutations of $P=(p_1,p_2,\cdots,p_N)$. By numerically solving
the sets of transcendental equations eq.(\ref {BAE}), the
quasimomentum $\left\{ k_j\right\} $ and thus the ground state
wavefunction can be determined exactly.

For the two-component Bose gas with SU(2) symmetry, it has been
proven that the ground states are $\left( N+1\right) $-fold
degenerate isospin `ferromagnetic' states \cite
{LiYQ,Gusj,Eisenberg,Yang}, which are symmetrical under permutation
of any two spins. Among the degenerate ground states, the fully
polarized state can be represented as
\begin{eqnarray}
\left| \Psi ^N\left( \frac N2,\frac N2\right) \right\rangle &=&\int d^N%
\mathbf{x}~\Psi \left( x_1,\cdots x_N\right) \times  \nonumber \\
&&\hat{\Psi}_{\uparrow }^{\dagger }(x_1)\hat{\Psi}_{\uparrow }^{\dagger
}(x_2)\cdots \hat{\Psi}_{\uparrow }^{\dagger }(x_N)\left| 0\right\rangle ,
\end{eqnarray}
where $\Psi \left( x_1,\cdots x_N\right) $ is given by the Bethe ansatz
ground-state wave function (\ref{wavefunction}). Other degenerate states can
be generated by applying the total lowering operator $\hat{S}^{-}$ to the
polarized state. For example, the total ground state wave function for the
degenerate ground state with $N_2$ spin-down particles (the state with $%
S=N/2 $ and $S_z=(N_1-N_2)/2$) can be expressed as
\begin{equation}
\left| \Psi ^N\left( \frac N2,\frac{N_1-N_2}2\right) \right\rangle =\left(
\hat{S}^{-}\right) ^{N_2}\left| \Psi ^N\left( \frac N2,\frac N2\right)
\right\rangle ,  \label{GS}
\end{equation}
where the total lowering spin operator $\hat{S}^{-}$ is defined as $\hat{S}%
^{-}=\frac 12\int dx\hat{\Psi}_{\downarrow }^{\dagger
}(x)\hat{\Psi}_{\uparrow }(x)$ .

\begin{figure}[tbp]
\includegraphics[width=3.5in]{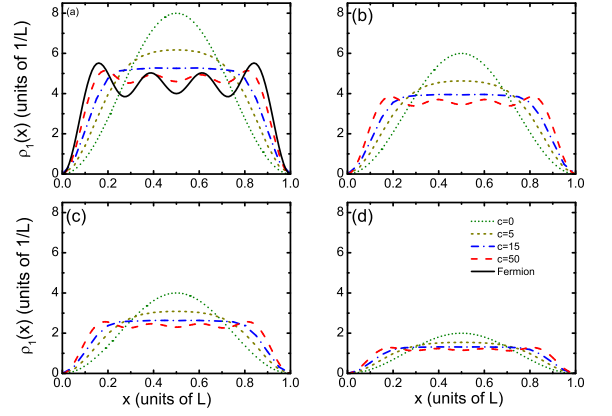}\newline
\caption{(color online) The ground-state density distribution of the
first
component for (a) $N_1=4, N_2=0$, (b)$N_1=3, N_2=1$, (c)$N_1=2, N_2=2$, (d)$%
N_1=1, N_2=3$, and $c=0, 5, 15, 50$. } \label{fig1}
\end{figure}

In terms of the ground state wave function $\Psi \left( x_1,\cdots
,x_N\right) $ the total density distribution $\rho _{\text{tot}%
}(x)=\sum_{\alpha =1,2}\rho _\alpha (x)$ can be expressed as
\[
\rho _{\text{tot}}(x)=\frac{N\int_0^Ldx_2\cdots dx_N\left| \Psi \left(
x,x_2,\cdots ,x_N\right) \right| ^2}{\int_0^Ldx_1\cdots dx_N\left| \Psi
\left( x_1,x_2,\cdots ,x_N\right) \right| ^2}.
\]
Here the ground-state density distribution of the $\alpha $-component is
given by
\[
\rho _\alpha (x)=\left\langle \Psi ^N\left( S,S_z\right) \right| \hat{\Psi}%
_\alpha ^{\dagger }(x)\hat{\Psi}_\alpha (x)\left| \Psi ^N\left( S,S_z\right)
\right\rangle .
\]
From the explicit form of the many body wavefunction, it is straightforward
to get the ground-state density distribution of the $\alpha $-component
which is found to fulfill a simple relation with the total density
distribution (see Appendix)
\begin{equation}
\rho _\alpha (x)=\frac{N_\alpha }N\rho _{\text{tot}}(x).  \label{ruo}
\end{equation}
We obtain unique total density profiles for all configurations with
the same total atom number $N$: $\left[ N_1,N_2\right] =\left[
N,0\right] ,\left[ N-1,1\right] ,\ldots \left[ 0,N\right] $, which
is also confirmed by the numerical exact diagonalization method in
the later evaluation. The conclusion (\ref{ruo}) is valid for the
integrable two-component boson system in the whole regime of
repulsive interaction. Moreover, we would like to indicate that the
conclusion (\ref{ruo}) is valid even in the presence of an external
confinement if the system has the total SU(2) symmetry, i.e., the
case with $U_1=U_2=U_{12}$ \cite{note}.  We thus recover the result
in Ref. \cite{Girardeau07} where only infinitely repulsive limit was
considered by a generalized Bose-Fermi mapping method.

In the following calculation, $L=1$ will be used through the paper. In Fig.
1 we display the the ground-state density distributions of the first
component for $N=4$ and $N_2=0,1,2,3$ for different interacting constants,
where we find similar crossover behavior as those of a single component Bose
gas. When the interaction is weak the density profiles show Gaussian-like
distribution and in the strongly interacting regime the density profiles
exhibit a shell structure with $N$ peaks for each component. In the
intermediate interacting regime the distribution show obvious evolution from
Bose distribution to Fermi distribution. According to eq.(\ref{ruo}), each
component takes the same density profiles and is normalized to the atom
number in the component. Therefore, all configurations with the same total
atom number share the same total density distribution. It is worth to note
the absence of the demixing in the integrable system, which is contrary to
prediction of a mean field approximation.

\begin{figure}[tbp]
\includegraphics[width=3.5in]{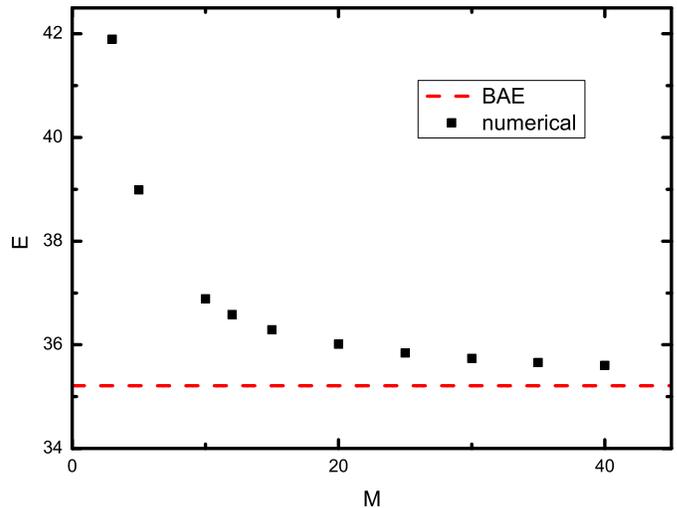}\newline
\caption{(color online) The ground state energy for $c=10$ and $N=4$
vs utilized orbital number. Dashed line: The exact result of Bethe
ansatz method; Scatters: Numerical diagonalization results. Units of
energy: $\hbar ^2/2mL^2$.} \label{fig2}
\end{figure}

\begin{figure}[tbp]
\includegraphics[width=3.5in]{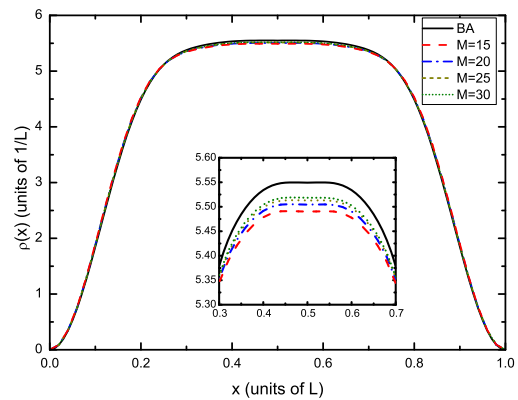}\newline
\caption{(color online) Density distribution of the ground state for $c=10$
and $N=4$. Inset: enlarged profiles in $x\in [0.3,0.7]$.}
\label{fig3}
\end{figure}

\section{Numerical Diagonalization Method}

Although the system is integrable for the situation of equal intra- and
inter-atomic interactions and some exact results can be obtained in this
case, we have to turn to the numerical method when the system deviates from
the integrable point. In fact it is very difficult to adjust the intra- and
inter-atomic interactions to be exactly the same in the realization of
experiment. For instance the scattering lengths and thus the effective 1D
interaction constants are known to be in the proportion $%
U_1:U_{12}:U_2=1.03:1:0.97$ in the two components of Bose gas composed of
internal states $\left|F=1,m_f=-1\right\rangle$ and $\left|F=2,m_f=1\right%
\rangle$ in $^{87}$Rb atoms \cite{scaleg}. In this situation Bethe ansatz
method is not applicable and we resort to the numerical exact
diagonalization method.

Let us first briefly review the numerical diagonalization method and then
investigate the ground state properties of the Bose-Bose mixture. The
normalized eigen wavefunction (orbital) of one particle in a hard wall takes
the form $\phi _i(x)=\sqrt{\frac 2L}\sin \left( \frac{ i\pi }Lx\right)$,
upon which the field operator $\Psi _\alpha (x)$ can be expanded as
\[
\Psi _\alpha (x)=\sum_{i=0}^\infty \phi _i(x)\hat{b}_{i\alpha }.
\]
The operator $\hat{b}_{i\alpha }^{\dagger }$ ($\hat{b}_{i\alpha }$) creates
(annihilates) one $\alpha $-component atom in the $i-$th orbital. As a
result the Hamiltonian (\ref{Ham}) is discretized as
\begin{eqnarray}
H &=&\sum_{\alpha =1,2}\left[ \sum_i\mu _i\hat{b}_{i\alpha }^{\dagger }\hat{b%
}_{i\alpha }+ {U_\alpha } \sum_{i,j,k,l}I_{i,j,k,l}\hat{b}_{i\alpha
}^{\dagger }\hat{b}_{j\alpha }^{\dagger }\hat{b}_{k\alpha }\hat{b}_{l\alpha
}\right]  \nonumber \\
&&+2 U_{12}\sum_{i,j,k,l}I_{i,j,k,l}\hat{b}_{i1}^{\dagger }\hat{b}%
_{j2}^{\dagger }\hat{b}_{k2}\hat{b}_{l1}
\end{eqnarray}
with $\mu _i=\left( i\pi \right) ^2$ ($i=1,2,3,...$) and the dimensionless
integrals $I_{ijkl}=\int_0^Ldx\phi _i(x)\phi _j(x)\phi _k(x)\phi _l(x)$. The
dimension of the Hilbert space is now $C_{N_1+M-1}^{N_1}\times
C_{N_2+M-1}^{N_2}$ if $N_1$ 1-component atoms and $N_2$ 2-component atoms
are populated on $M$ orbitals. Then the ground state $\left| \text{GS}%
\right\rangle$ can be obtained after diagonalizing the Hamiltonian in the
Hilbert space spanned by the one-particle eigenstates. In order to assure
the precision of evaluation sufficient orbitals should be considered
particularly for the systems in strongly interacting regime. 
The total density distribution is given by
\[
\rho _{\text{tot}}(x)=\sum_{\alpha =1}^2\rho _\alpha (x)
\]
with the density distribution of $\alpha$-component
\begin{eqnarray}
\rho _\alpha (x) &=&\left\langle \text{GS}\right| {\Psi }_\alpha ^{\dagger
}(x){\ \Psi }_\alpha (x)\left| \text{GS}\right\rangle  \nonumber \\
&=&\sum_{i,j}\phi _i^{*}(x)\phi _j(x)\left\langle \text{GS}\right| \hat{b}%
_{i\alpha }^{\dagger }\hat{b}_{j\alpha }\left| \text{GS}\right\rangle .
\end{eqnarray}
In terms of the ground state wave function the reduced one-body density
matrix for each component and the two body correlation of intra- and
inter-species atom can be formulated as
\begin{eqnarray}
\rho _\alpha (x,x^{\prime }) &=&\left\langle \text{GS}\right| {\Psi }_\alpha
^{\dagger }(x){\ \Psi }_\alpha (x^{\prime })\left| \text{GS}\right\rangle
\nonumber \\
&=&\sum_{i,j}\phi _i^{*}(x)\phi _j(x^{\prime })\left\langle \text{GS}\right|
\hat{b}_{i\alpha }^{\dagger }\hat{b}_{j\alpha }\left| \text{GS}\right\rangle
\end{eqnarray}
and
\begin{eqnarray}
& & \rho _{\alpha \beta}(x,x^{\prime })  \nonumber \\
&=&\left\langle \text{GS}\right| {\Psi }_{\alpha}^{\dagger }(x){\ \Psi }%
_{\alpha}(x){\Psi }_{\beta}^{\dagger }(x^{\prime }){\ \Psi }%
_{\beta}(x^{\prime })\left| \text{GS} \right\rangle \\
&=&\sum_{i,j,k,l}\phi _i^{*}(x)\phi _j(x)\phi _k^{*}(x^{\prime })\phi
_l(x^{\prime })\left\langle \text{GS}\right| \hat{b}_{i\alpha}^{\dagger }%
\hat{b} _{j\alpha}\hat{b}_{k\beta}^{\dagger }\hat{b}_{l\beta}\left| \text{GS}%
\right\rangle .  \nonumber
\end{eqnarray}
The momentum distribution is simply the Fourier transformation of $\rho
_\alpha (x,x^{\prime })$,
\begin{equation}
n_\alpha \left( k\right) =\frac 1{2\pi }\int_0^Ldx\int_0^Ldx^{\prime }\rho
_\alpha (x,x^{\prime })e^{-ik\left( x-x^{\prime }\right) }.
\end{equation}

In order to test the accuracy of our numerical code, we compare the
numerical result with that from Bethe ansatz method on calculating the
ground state energy and total density profiles of two components Bose gas
with $N=4$. The results are shown in Fig. 2 and Fig. 3 for an intermediate
interaction constant $c=10$. The ground state energy is shown to
asymptotically approach the BA result $E/N=35.22$ if sufficient orbitals are
taken into account. For instance, we have the ground state energy $E/N=35.60$
for $M=40$, the deviation of which is already within $1\%$. The density
profiles calculated for $M=15$ can match the BA result very well. In the
following calculation, the orbital number is taken as $M=20$ ($M=15$) for $%
N=4$ $(N=5)$ and the reduced Hilbert space is typically composed of $10^4$
basis states with a corresponding energy cutoff $(M \pi )^2=3947.84$ $%
(2220.66)$.

\begin{figure}[tbp]
\includegraphics[width=3.8in]{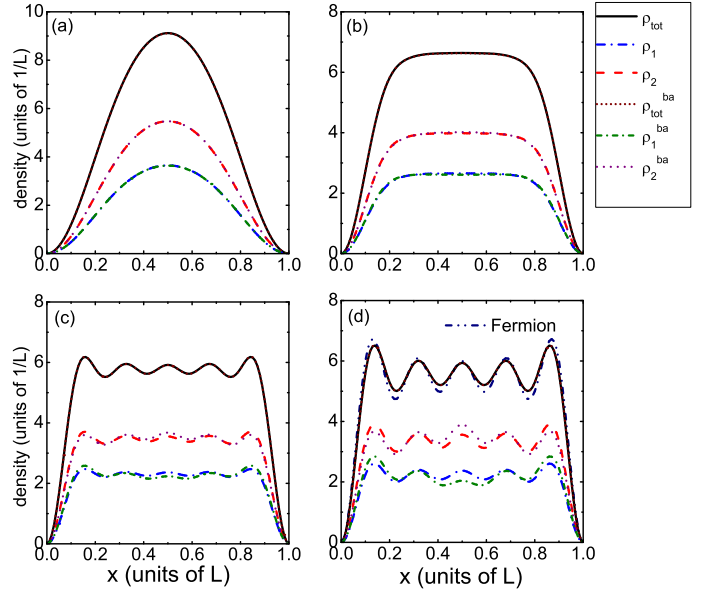}\newline
\caption{(color online) Density distribution of the ground state for $%
N_1=2,N_2=3$ and $U=1.0(a),10.0(b),40.0(c),80.0(d)$. $\rho _\alpha ^{\text{BA%
}} (\alpha =1,2,\text {tot})$: Bethe-Ansatz result for equal intra- and
inter-atomic interactions ($U_1=U_2=U_{12}=U$); $\rho _\alpha (\alpha =1,2,
\text {tot})$: Numerical result for unequal intra- and inter-atomic
interactions ($U_1:U_{12}:U_2=1.03:1:0.97$).}
\label{fig4}
\end{figure}

Using the numerical method the density profiles of each component can be
obtained even for unequal atom numbers and $SU(2)$ symmetry broken atomic
interaction constants, i.e. $U_1 \neq U_2\neq U_{12}$. In Fig 4. we display
both the total density distribution and the that of each component in the
full interacting regime for equal and unequal intra- and inter-atomic
interactions in the case of $N_1=2,N_2=3$. In the situation of unique
interaction constant, two components display the same density distribution
in the full interacting regime, i.e., $\rho _\alpha(x)=\frac{N_\alpha}N\rho
_{\text{tot}}(x)$. The density profiles show evolution from Bose to Fermi
distribution with the increase of atomic interaction. In Fig. 4d we compare
the distribution of the system of finite strong interaction with the
distribution of TG gas, which is obtained using the Bose-Fermi mapping. It
turns out that even if the interaction is finite the result from Bose-Fermi
mapping can describe the system very well. Generally the ground state energy
and the density profiles of two-component Bose gas do not show distinct
difference from its single component counterpart. 
For unequal intra- and inter-atomic interacting constants as in the
experiment ($U_1:U_{12}:U_2=1.03:1:0.97$), the density profiles do not
change drastically comparing with those of integrable system even in the
strongly interacting regime. Particularly in the weakly interacting regime,
the exact solution of integrable system provides a trustable description of
the real experimental system because of the relatively small asymmetry of
the intra- and inter-species interacting constants .
\begin{figure}[tbp]
\includegraphics[width=3.5in]{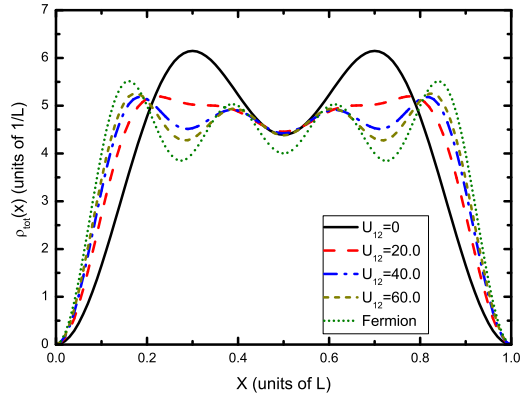}\newline
\caption{(color online) Density distribution of the ground state for $%
N_1=N_2=2$ and $U_1=U_2=60$.}
\label{fig5}
\end{figure}

\section{Interaction between two TG gases}

We have shown that how the fermionization crossover for the one-component
Bose gas extends to a two-component mixture in the whole repulsive regime
with almost equal intra- and inter-atomic interactions. The numerical
diagonalization method can be used to investigate two components of Bose
gases even when there are great difference between the intra- and
inter-atomic interactions. Now we focus on the crossover induced by the
inter-atomic interaction constants. We start with the case with $U_{12}=0$
and $U_1=U_2=60$, where each component is an independent TG gas, and
increase the inter-atomic interaction $U_{12}$ to see how the composite
fermionization crossover happens for two initially fermionized components.
In Fig. 5 the total density profiles are given for $N_1=N_2=2$ and $%
U_1=U_2=60$. In this situation the distributions of each component
match each other and they are one half of the total distribution
because of equal atom numbers in two components. When the
inter-atomic interaction disappear the system is composed of two
isolated TG gases that display $N_\alpha$ peaks. With the increase
of inter-atomic interaction the density profiles become flatter with
more peaks appearing. In the strongly interacting limit the shell
structure with $N$ peaks display, which is the same as the density
profiles of single component of TG gas of $N$ atoms. For the 1D
Bose-Bose mixtures under harmonic confinement, such a
composite-fermionization crossover has been observed as the
interspecies coupling strength is varied to the limit of infinite
repulsion \cite{Zoellner08,Hao08}.

\begin{figure}[tbp]
\includegraphics[width=3.3in]{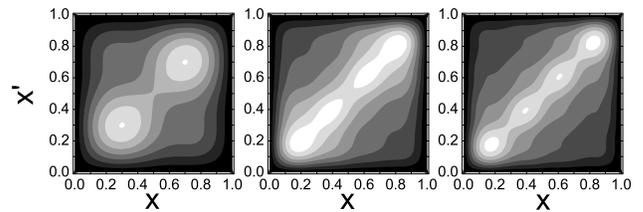}\newline
\caption{The reduced one body density matrix for each component of
the ground state for $N_1=N_2=2$ and $U_1=U_2=60.0$.
$U_{12}=0.0,30.0,60.0$ (from left to right.) Units of length: $ L$.}
\label{fig6}
\end{figure}
\begin{figure}[tbp]
\includegraphics[width=3.5in]{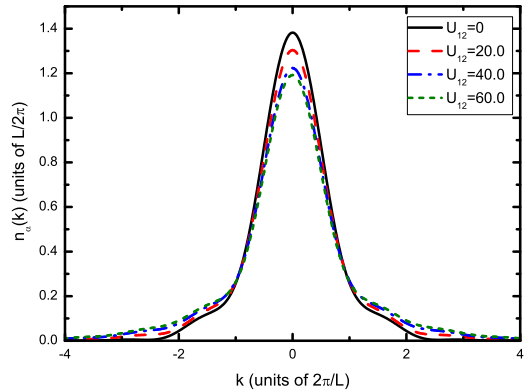}\newline
\caption{(color online) Momentum distribution of the ground state for $%
N_1=N_2=2$, $U_1=U_2=60.0.$}
\label{fig7}
\end{figure}
In Fig. 6 we show the reduced one body density matrix for each component,
which means the probability that two successive measurements, one
immediately following the other, will find the same component particle at
the point $x$ and $x^{\prime }$, respectively. We notice that for all
interacting strengthes there exists a strong enhancement of the diagonal
contribution $\rho_\alpha (x,x^{\prime })$ along the line $x=x^{\prime }$.
The identical momentum distributions for the two components are shown in
Fig. 7. For all inter-atomic interaction strengths, Bose atoms accumulate in
the central regime close to zero momentum and the distributions decrease
rapidly for large momentum. For strong inter-component interaction, the
momentum distribution becomes broader and broader with the $k=0$ peak
diminishing.

\begin{figure}[tbp]
\includegraphics[width=3.3in]{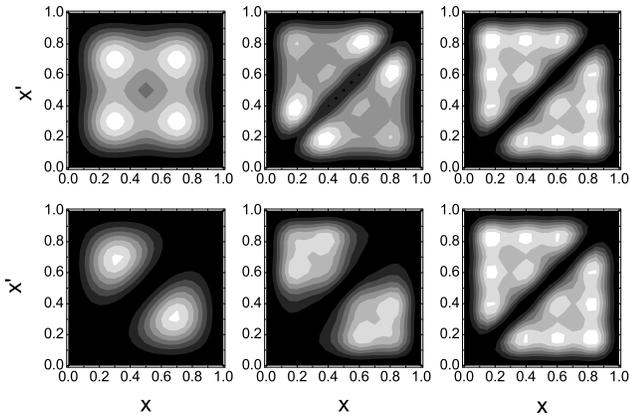}\newline
\caption{Two-body correlation of intra- and inter-component of the
ground state for $N_1=N_2=2$ and $U_1=U_2=60.0$. Top: correlation
between two atoms in different components; bottom: correlation
between two atoms in the same component. $U_{12}=0.0,30.0,60.0$
(from left to right.) Units of length: $ L$.} \label{fig8}
\end{figure}

It is also interesting to study the density-density correlation functions,
which denote the probability that one measurement will find a atom at the
point $x$ and the other one at the point $x^{\prime }$. In Fig. 8 we display
the intra- and inter-species correlations between two atoms. At $U_{12}=0.0$
we have two uncorrelated TG gases and thus $\left\langle \hat{\rho}_1(x)
\hat{\rho}_2 (x^{\prime}) \right\rangle = \left\langle \hat{\rho}_1(x)
\right\rangle \left\langle \hat{\rho}_2 (x^{\prime}) \right\rangle $. With
the increase of the inter-atomic interaction two components will try to
avoid each other and it becomes more unlikely that one will find two atoms
in different components at the same position. The intra-species correlation
is always small in all cases because of the strong intra-atomic interactions
in TG limit.

\section{Summary}

In conclusion we have investigated the ground state of two-component Bose
gas with Bethe ansatz method and numerical diagonalization method. It turns
out that the numerical result describe the ground state of the system
quantitatively well. When intra- and inter-atomic interaction are same ($%
U_1=U_2=U_{12}$), the two-components Bose gas is integrable and the ground
state wavefunction can be obtained exactly. The ground state energy and
total density distribution are same for all configurations with same total
atom number. With the increase of the interaction the total density
distribution show evolution from a Gauss-like Bose distribution (one peak)
to a shell structure of nointeracting spinless Fermions ($N$ peaks). The
distribution of each component is $N_\alpha/N$ of the total density
distribution. If the interaction constants deviate the integrable point a
little, which is the real situation in experimental, the Bose mixture shows
almost same behaviors as the integrable system. In addition we investigate
the effect induced by the inter-atomic interaction constants for two TG
gases with the numerical diagonalization method. It turns out that with the
increase of interspecies interaction the system shows composite
fermionization crossover.

\begin{acknowledgments}
This work was supported by NSF of China under Grants No. 10574150
and No. 10774095, the 973 Program under Grants No. 2006CB921102 and
No. 2006CB921300, and National Program for Basic Research of MOST,
China. S.C. is also supported by programs of Chinese Academy of
Sciences. He would like to thank S. J. Gu for helpful discussion.
\end{acknowledgments}

\appendix
\section{}

The total ground state wave function with $N_2$ spin-down particles
(the state with $S=N/2$ and $S_z=(N_1-N_2)/2$) can be generated by
applying the total lowering operator $\hat{S}^{-}=\frac 12\int
dx\hat{\Psi}_{\downarrow }^{\dagger }(x)\hat{\Psi}_{\uparrow }(x)$
to the polarized state according to eq.(7) and its normalized
formula shall be expressed as
\begin{eqnarray}
&&\left| \Psi ^N\left( \frac N2,\frac{N_1-N_2}2\right) \right\rangle  \\
&=&\frac 1{\sqrt{c}}\int d^N\mathbf{x}\Psi \left( x_1,x_2,\cdots
,x_N\right)
\nonumber \\
&&\times \hat{\Psi}_{\downarrow }^{\dagger }(x_1)\cdots \hat{\Psi}%
_{\downarrow }^{\dagger }(x_{N_2})\hat{\Psi}_{\uparrow }^{\dagger
}(x_{N_2+1})\cdots \hat{\Psi}_{\uparrow }^{\dagger }(x_N)\left|
0\right\rangle   \nonumber
\end{eqnarray}
with the normalized constant
\begin{eqnarray*}
c &=&N_1!N_2!\int dx_1\cdots dx_N \\
&&\times \Psi \left( x_1,x_2,\cdots ,x_N\right) ^{*}\Psi \left(
x_1,x_2,\cdots ,x_N\right) .
\end{eqnarray*}
The density distribution of $\alpha $th component can be expressed
as
\begin{eqnarray*}
&&\rho _\alpha (x)=\left\langle \Psi ^N\left( S,S_z\right) \right| \hat{\Psi}%
_\alpha ^{\dagger }(x)\hat{\Psi}_\alpha (x)\left| \Psi ^N\left(
S,S_z\right)
\right\rangle  \\
&=&\frac 1c\int d^N\mathbf{y}d^N\mathbf{x}\Psi \left( y_1,\cdots
y_N\right)
^{*}\Psi \left( x_1,\cdots x_N\right)  \\
&&\times \left\langle 0\right| \hat{\Psi}_{\uparrow }(y_N)\cdots \hat{\Psi}%
_{\uparrow }(y_{N_2+1})\hat{\Psi}_{\downarrow }(y_{N_2})\cdots \hat{\Psi}%
_{\downarrow }(y_1)\hat{\Psi}_\alpha ^{\dagger }(x) \\
&&\times \hat{\Psi}_\alpha (x)\hat{\Psi}_{\downarrow }^{\dagger
}(x_1)\cdots \hat{\Psi}_{\downarrow }^{\dagger
}(x_{N_2})\hat{\Psi}_{\uparrow }^{\dagger }(x_{N_2+1})\cdots
\hat{\Psi}_{\uparrow }^{\dagger }(x_N)\left| 0\right\rangle
\end{eqnarray*}
For the first component the above formulation can be evaluated as
follows
\begin{eqnarray*}
&&\rho _1(x)=\left\langle \Psi ^N\left( S,S_z\right) \right| \hat{\Psi}%
_{\uparrow }^{\dagger }(x)\hat{\Psi}_{\uparrow }(x)\left| \Psi
^N\left(
S,S_z\right) \right\rangle  \\
&=&\frac 1c\int d^N\mathbf{y}d^N\mathbf{x}\Psi \left( y_1,\cdots
y_N\right)
^{*}\Psi \left( x_1,\cdots x_N\right)  \\
&&\times \left\langle 0\right| \hat{\Psi}_{\uparrow }(y_N)\cdots \hat{\Psi}%
_{\uparrow }(y_{N_2+1})\hat{\Psi}_{\uparrow }^{\dagger }(x) \\
&&\times \hat{\Psi}_{\uparrow }(x)\hat{\Psi}_{\uparrow }^{\dagger
}(x_{N_2+1})\cdots \hat{\Psi}_{\uparrow }^{\dagger }(x_N) \\
&&\times \hat{\Psi}_{\downarrow }(y_{N_2})\cdots
\hat{\Psi}_{\downarrow }(y_1)\hat{\Psi}_{\downarrow }^{\dagger
}(x_1)\cdots \hat{\Psi}_{\downarrow
}^{\dagger }(x_{N_2})\left| 0\right\rangle  \\
&=&\frac 1cN_1N_1!N_2!\int dx_2\cdots dx_N \\
&&\times \Psi \left( x,x_2,\cdots ,x_N\right) ^{*}\Psi \left(
x,x_2,\cdots ,x_N\right) .
\end{eqnarray*}
Similarly, for the second component we have
\begin{eqnarray*}
&&\rho _2(x)=\left\langle \Psi ^N\left( S,S_z\right) \right| \hat{\Psi}%
_{\downarrow }^{\dagger }(x)\hat{\Psi}_{\downarrow }(x)\left| \Psi
^N\left(
S,S_z\right) \right\rangle  \\
&=&\frac 1c\int d^N\mathbf{y}d^N\mathbf{x}\Psi \left( y_1,\cdots
y_N\right)
^{*}\Psi \left( x_1,\cdots x_N\right)  \\
&&\times \left\langle 0\right| \hat{\Psi}_{\uparrow }(y_N)\cdots \hat{\Psi}%
_{\uparrow }(y_{N_2+1})\hat{\Psi}_{\uparrow }^{\dagger
}(x_{N_2+1})\cdots
\hat{\Psi}_{\uparrow }^{\dagger }(x_N) \\
&&\times \hat{\Psi}_{\downarrow }(y_{N_2})\cdots
\hat{\Psi}_{\downarrow
}(y_1)\hat{\Psi}_{\downarrow }^{\dagger }(x)\hat{\Psi}_{\downarrow }(x)\hat{%
\Psi}_{\downarrow }^{\dagger }(x_1)\cdots \hat{\Psi}_{\downarrow
}^{\dagger
}(x_{N_2})\left| 0\right\rangle  \\
&=&\frac 1cN_2N_1!N_2!\int dx_2\cdots dx_N \\
&&\times \Psi \left( x,x_2,\cdots ,x_N\right) ^{*}\Psi \left(
x,x_2,\cdots ,x_N\right) .
\end{eqnarray*}
Thus the total ground-state density distribution is given by
\begin{eqnarray*}
&&\rho _{\text{tot}}(x)=\rho _1(x)+\rho _2(x) \\
&=&\frac{N\int dx_2\cdots dx_N\Psi \left( x,x_2,\cdots ,x_N\right)
^{*}\Psi \left( x,x_2,\cdots ,x_N\right) }{\int dx_1\cdots dx_N\Psi
\left( x_1,x_2,\cdots ,x_N\right) ^{*}\Psi \left( x_1,x_2,\cdots
,x_N\right) }
\end{eqnarray*}
with the density distribution of $\alpha $-component
\[
\rho _\alpha (x)=\frac{N_\alpha }N\rho _{\text{tot}}(x).
\]


\begin{thebibliography}{99}
\bibitem{note} This is true because the conclusion that the ground state
for a 2-component boson system with the total SU(2) symmetry is
$(N+1)$-foldly degenerate is indepedent of the external potential.
Therefore, as long as the total SU(2) symmetry is not broken, one
can construct the degenerate states according to Eq. (7). The
derivation of Eq. (8) in the appendix does not depend on the
concrete form of $\Psi(x_1,\cdots,x_N)$ except the property of its
exchange symmetry. However when there exists an external potential,
for example, a harmonic potential, one can not calculate the
wavefunction exactly since the system is no longer integrable.
\end{thebibliography}
\end{document}